 \documentclass[aps,pre,reprint]{revtex4-1}
 
\usepackage{amsmath}    % need for subequations
\usepackage{graphicx}   % need for figures
\usepackage{verbatim}   % useful for program listings
\usepackage{color}      % use if color is used in text
\usepackage{subfigure}  % use for side-by-side figures
\usepackage{hyperref}   % use for hypertext links, including those to external documents and URLs
\begin{document}
% \addtocounter{MaxMatrixCols}{7}
\title{Reduction of electron heating by magnetizing ultracold neutral plasma}
\author{Sanat Kumar Tiwari}
\email{sanat-tiwari@uiowa.edu}
\affiliation{Department of Physics and Astronomy, University of Iowa, Iowa City, Iowa 52242, USA}
\author{Scott D.~Baalrud}
\affiliation{Department of Physics and Astronomy, University of Iowa, Iowa City, Iowa 52242, USA}
\date{\today}%\emph{•}
%{for revtex  maketitle should be written just here}
%\maketitle 
\begin{abstract} 
Electron heating in an ultracold neutral plasma is modeled using classical molecular dynamics simulations in the presence of an externally applied magnetic field. A sufficiently strong magnetic field is found to reduce disorder induced heating and three body recombination heating of electrons by constraining electron motion, and therefore heating, to the single dimension aligned with the magnetic field.  A strong and long-lasting temperature anisotropy develops, and the overall kinetic electron temperature is effectively reduced by a factor of three. These results suggest that experiments may increase the effective electron coupling strength using an applied magnetic field. 
\end{abstract}
% for revtex4 here maketitle should be written
%\pacs{} 
\maketitle 
\section{Introduction}
\label{intro}
\paragraph*{}
Ultracold neutral plasma (UCP) experiments provide an excellent testbed for theories of transport properties of moderate to strongly coupled plasmas because they can be precisely diagnosed~\cite{strickler_16}. Such measurements are interesting from a fundamental physics perspective because they access exotic regimes of plasma physics. Validating theory also contributes to other areas of science, such as inertial confinement fusion \cite{Hu_prl_10,drake2006high} and astrophysics \cite{muchmore_84,lben_85}, where moderate and strongly coupled plasmas are encountered in systems that are comparatively difficult to diagnose. In a typical UCP experiment, the coupling strength of electrons and ions are limited to $\Gamma_e \lesssim 0.1$ and $\Gamma_i \lesssim 5$, respectively, due to disorder induced heating (DIH) and three-body recombination (TBR) heating processes \cite{Killian07,bannasch_13}. Here, 
%%%%%%%%%%%%%%%%%%%%%%
\begin{equation}
\Gamma_s = \frac{Z_s^2e^2/a_s}{k_BT_s}
\end{equation}
%%%%%%%%%%%%%%%%%%%%%%
is the ratio of the average Coulomb potential energy of species `s' at the species interparticle spacing, $a_s = (3/4\pi n_s)^{1/3}$, to the average kinetic energy $k_BT_s$, and $Z_s$ and $n_s$ are the charge (in electron units) and number density of species `s', respectively.  If either the electron coupling strength, ion coupling strength, or both, could be increased, UCPs could reach physics regimes that are more interesting from a fundamental physics standpoint and which more directly resemble conditions in many of the other applications of interest. 

In this paper, classical molecular dynamics (MD) simulations are used to show that a strong external magnetic field can reduce electron heating, leading to a larger electron coupling strength.  The electron temperature is reduced because electron motion is restricted to nearly one-dimension, parallel to the direction of the applied magnetic field $\bf{B}$.  
In this paper, electron heating is studied for times up to 10 $\omega_{pe}^{-1}$, where $ \omega_{pe} =\sqrt{4\pi e^2 n_e/m_e}$ is the electron plasma frequency. Over  this duration, neither plasma species  acquires thermodynamic equilibrium. The notion of temperature is represented by the  ``kinetic temperature" defined as:
%%%%%%%%%%%%%%%%%%%%%%
\begin{subequations}
\begin{equation}
 k_B T_s^\parallel = \frac{2}{N_s}\sum_{i=1}^{N_s} \frac{1}{2} m_s |\dot{r}_x|^2_{si}
 \label{temp_prl}
\end{equation}
\begin{equation}
 k_B T_s^\perp = \frac{1}{N_s}\sum_{i=1}^{N_s} \frac{1}{2} m_s (|\dot{r}_y|^2_{si}+|\dot{r}_z|^2_{si})
 \label{temp_pl}
\end{equation}
\begin{equation}
T_s = \frac{1}{3}(T_s^\parallel + 2T_s^\perp)  .
\label{temp_tot}
\end{equation}
\end{subequations}
%%%%%%%%%%%%%%%%%%%%%%
Here, $T_s^\parallel$ and $T_s^\perp$ are the kinetic temperatures for species `s' in the parallel (x direction) and the perpendicular direction to the applied magnetic field respectively. $N_s$ and $m_s$ are the total number of particles and the mass of single particle of species `s', respectively. 
Heating occurs predominantly in the parallel direction, so the total kinetic temperature is observed to be approximately one-third of the value obtained in the absence of $\bf{B}$.  A large temperature anisotropy is established as a result. Eventually this anisotropy relaxes, and heating also occurs in the perpendicular directions. However, if the magnetic field is strong enough, the relaxation can be dramatically suppressed \cite{Glinsky_pof_92,Beck_92,dubin_05,Anderegg_09,ott_pre_17a}, 
%. If the field is strong enough, the temperature anisotropy relaxation may 
and may be delayed long enough that measurements could be made before heating occurs in the perpendicular directions. 
This may increase the effective Coulomb coupling strength, $\Gamma$ as measured in terms of the kinetic temperature.

\paragraph*{}
In experiments, creation of UCP is followed by DIH on a timescale characterized by $\omega_{pe}^{-1} \sim$ (nanoseconds) \cite{kuzmin_pop_02,lyon_2011,maxson_13} and later by TBR heating \cite{fletcher_prl_07}. Ions are also influenced by DIH, but on a longer timescale characterized by the ion plasma period ($\omega_{pi}^{-1}$) \cite{cheny_04,McQuillen_15}.  Previous and ongoing efforts to increase coupling strength largely focus on ions, including laser cooling \cite{langin_cooling_17,shaparev_15},
introducing initial spatial correlation using blockaded Rydberg gases \cite{bannasch_13} or Fermi gases \cite{murillo_01}.  
or by trapping the gas in an optical lattice \cite{guo_10,murphy_16}. We focus on electrons because they are more easily magnetized than ions. 

\paragraph*{}
Experiments have shown that a magnetic field of a modest magnitude ($\sim 50$ G) can substantially reduce transverse expansion of an ultracold plasma \cite{zhang_prl_08}. Previous theoretical studies have also suggested that a strong magnetic field can substantially reduce the rate of TBR in a weakly coupled plasma, and therefore the heating associated with it~\cite{Glinsky_91,Robicheaux_06}. The magnetization strength of a species can be quantified by the parameter $\beta_s = \omega_{cs}/\omega_{ps}$, which is  the ratio of the gyrofrequency ($\omega_{cs} = q_s B/m_s$)  to the plasma frequency.
Here, we focus on the regime where electrons are strongly magnetized $\beta_e \gg 1$. 
In this regime electrons are constrained to small gyro-orbits, which reduces their collision rate significantly in the transverse direction. 
This suppresses DIH in the transverse direction. 
The weakest magnetic field at which any influence is observed is $\beta_e \simeq 1$. 
At the density \textcolor{black}{$n \simeq 10^8$ cm$^{-3}$} relevant to UCPs, $\beta_e = 1$ if \textcolor{black}{$B=32$ G}. 
A much more substantial reduction in transverse DIH is observed for $\beta_e \gtrsim 10$ (corresponding to \textcolor{black}{$B \gtrsim 320$ G}). 
Previously, Glinsky et al \cite{Glinsky_pof_92}, Dubin et al \cite{dubin_05,Anderegg_09} and Ott et al \cite{ott_pre_17a} have shown that the long lived temperature anisotropy can be maintained for $\beta>1$ in weakly, moderately and strongly coupled plasmas, respectively.
Recently, Baalrud et al has provided the temperature anisotropy relaxation rates ranging from weak to strong coupling strength regime and for magnetic field strength $\beta$ ranging from 0  to 100 using MD simulations \cite{baalrud_17}.
Based on their data, it is expected that a magnetization strength of $\beta_e \gtrsim 100$ (corresponding to \textcolor{black}{$B\gtrsim 3200$ G}) may be sufficient to delay relaxation of the electron temperature anisotropy long enough ($\sim 10^6 \omega_{pe}^{-1}$) to make measurements over the duration of a typical experiment ($\sim 100 \mu$s) \cite{McQuillen_15,Killian07}. 
%%%%%%%%%%%%%%%%%%
\begin{figure}
 \includegraphics[width = 8cm]{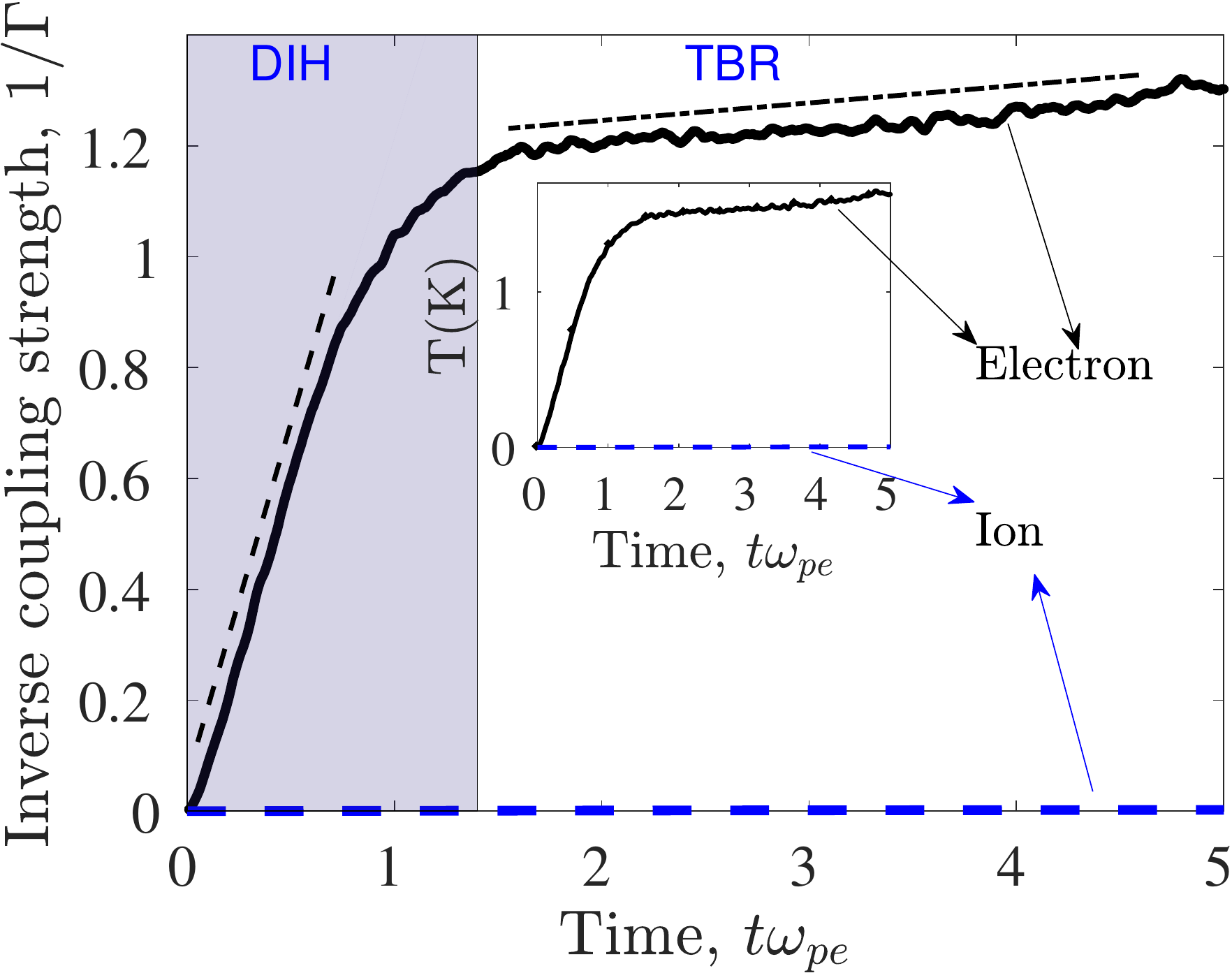}
 \caption{Temperature evolution measured in terms of $\Gamma^{-1}$. Simulations were carried out with initial temperatures $T_e = T_i = 0$ and a homogeneous random distribution of electrons and ions. The repulsive core scale length was $\alpha = 0.01$  and the mass ratio $m_i/m_e = 1.62 \times 10^5$ (Strontium).}
 \label{dih_ei}
\end{figure}
%%%%%%%%%%%%%%%%%%

In the strong field regime, we also observe a reduced rate of heating due to TBR. Like DIH, the reduction is observed to be a factor of three in the strong field regime ($\beta_e \gtrsim 20$). Previous theoretical work has suggested that the rate of TBR may be suppressed by a factor of 10 in a strong magnetic field at weak coupling. Our work 
%provides the first study of the effect of a magnetic field on the associated heating rate in the
extends this to the moderately coupled regime.

\paragraph*{}
The paper is organized as follows. Section \ref{sim_mod} provides details of the simulation model.  Section \ref{dih_nomag} discusses heating processes in  unmagnetized ultracold plasma, and this is extended to include strong magnetization in Sec.~\ref{dih_mag}. Finally, we summarize in Sec.~\ref{summ}.

\section{Simulation model}
\label{sim_mod}
\paragraph*{}
%%%%%%%%%%%%%%%%%%
\begin{figure}%[!htb]
 \includegraphics[width = 8cm]{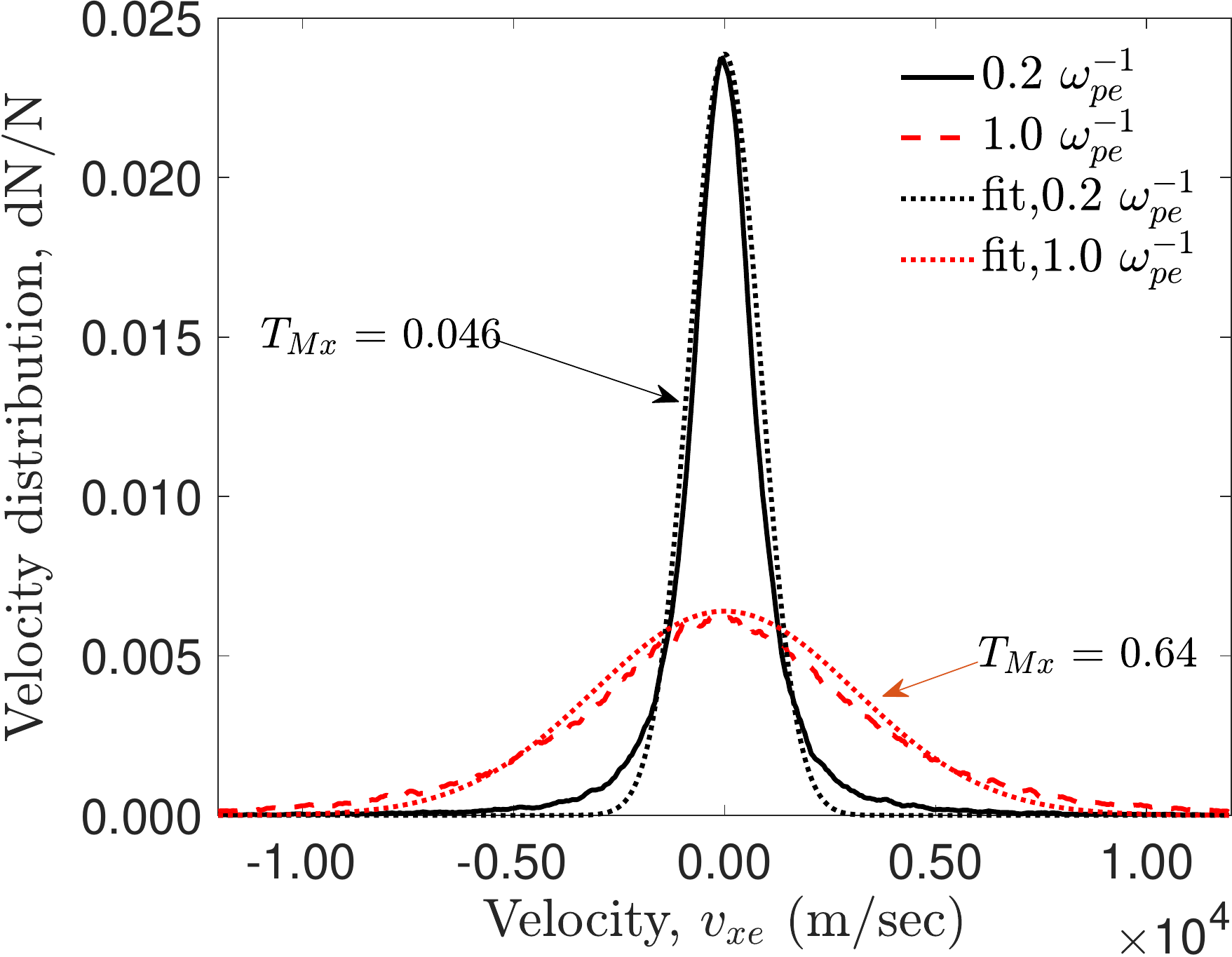}
\caption{Electron velocity distribution at 0.2 $\omega_{pe}^{-1}$ (black line) and 1  $\omega_{pe}^{-1}$ (red line)  during the DIH process. 
Dotted lines are the Maxwellian fit corresponding to $T_{Mx}$. dN is the number of particles 
within each velocity bin and N is total number of particles. The width of each velocity bin is from $v_x$ to $v_x + 50$~m/s.} 
 \label{dih_edistr}
\end{figure}
%%%%%%%%%%%%%%%%%%
Non-equilibrium classical MD simulations \cite{kuksin_05} were carried out using the open source code LAMMPS~\cite{Plimpton1995}.   The simulation model involved a cubic box with periodic boundary conditions in each of the three dimensions. Equal numbers of electrons and ions ($10^4$ particles of each species) were introduced at random positions, so as to avoid any initial spatial pre-correlation amongst the charged particles that would have a suppressing effect on DIH~\cite{bannasch_13, murphy_16}.  Each simulation was carried out with three different initial random configurations and the presented results are a mean of these  independent runs. 
Each species was initialized with zero temperature (i.e. all particles had zero initial velocity). Simulations were also carried out with a very small but finite initial plasma temperature (corresponding to  $\Gamma \approx  200$) and the results were confirmed to be the same as those with zero initial temperature.
The results shown focus on plasma with singly charged strontium ions ($m_{i} = 1.62 \times 10^5 m_e$) to model a common experimental choice. Cases with different ion to electron mass ratio will be stated explicitly.  At such a high mass ratio, ions were approximately stationary over the timescale of several $\omega_{pe}^{-1}$ that is the focus of the present study. It was confirmed that similar results were obtained over this time interval if ions were kept stationary at their initial positions. 

An electron-ion plasma was modeled taking like-charged particles to interact through the repulsive Coulomb potential and particles with opposite charge to interact through the attractive Coulomb potential along with a repulsive core 
%%%%%%%%%%%%%%%%%%%
\begin{subequations}
\label{eq:pots}
 \begin{eqnarray}
 && v_{ee} = v_{ii} = \frac{e^2}{r}  \\
 && v_{ei} = -\frac{e^2}{r} \left[ 1 - \exp{\left(-\frac{r^2}{(\alpha a)^2}\right)} \right] .
 \label{pot_md}
 \end{eqnarray}
 \end{subequations}
 %%%%%%%%%%%%%%%%%%%
Here, $r$ is the separation between two charged particles, $e$ is the elementary charge,  and $\alpha$ is an adjustable parameter that sets the e-i repulsion length scale. The repulsive core was included to avoid Coulomb collapse between oppositely charged point particles. For the purpose of this work, $\alpha$ was chosen to be sufficiently small that it did not influence the results in any way. It is a purely numerical necessity to avoid the rare occurrence of the very close interactions between electrons and ions, which is necessary to prevent because arbitrarily close interactions can not be resolved by a finite timestep integration, 
leading to energy conservation issues.

\paragraph*{•} 
Recently, Tiwari et al. used this model to calculate thermodynamic state variables in recombining ultracold plasmas ~\cite{sanat_thermo_17}.
It was found that the magnitude of $\alpha$ controls the concentration of free charges compared to classical Rydberg states because it determines the depth of the potential well formed between oppositely charged particles.  By distinguishing the free and classically bound populations, it was shown that the excess pressure and excess internal energy of the free charges becomes independent of the choice of repulsive core length scale length when it is sufficiently short-ranged. This enabled simulations of thermodynamic properties of an ultracold plasma at fixed conditions. Here, it is applied to study dynamic properties on short timescales. This is a simpler situation because few bound states are formed on the short timescales of interest. 

The system was evolved according to the equation of motion 
%%%%%%%%%%%%%%%%%%%
\begin{equation}
m_j \ddot {\bf r}_j = {\bf F}_j + m_j \omega_{cj} \dot{\bf r}_j \times {\bf e}_x
\label{ucold_dyn}
\end{equation}
%%%%%%%%%%%%%%%%%%%
where $\mathit{j}$ is the index of any particle in the system ($j = 1 \ldots N$), 
${\bf F}_j = -\sum_{i=1}^{N-1} \nabla {v}_{ij}$ is the total force on $\mathit{j}^{th}$ particle due to interaction with all other particles, and ${\bf B} = \mathrm{B} {\bf e}_x$ is the magnetic field.  A modified Velocity-Verlet scheme (as described by Spreiter el al. \cite{spreiter_99})  was  used to evolve the equation of motion \cite{Tim_lammps}. The time step was chosen to resolve the gyromotion, electron plasma frequency and the orbits of electron-ion interactions. 
For $\beta \leq 10$, it was chosen to be $\Delta t = 10^{-3} \omega_{pe}^{-1}$, and for $\beta >10$ to be $\Delta t = 10^{-2} \omega_{ce}^{-1}$ 
The total energy was conserved to within 0.01 $\%$ for a time span of 10 $\omega_{pe}^{-1}$. 

Ultracold plasma evolution has been modeled using classical molecular dynamics with a soft repulsive core potential by Kuzmin et al \cite{kuzmin_pop_02,kuzmin_prl_02}, with a hybrid molecular dynamics model by Pohl et al \cite{Pohl_04} and recently by Isaev et al. to study the heating and diffusion processes in the presence of a homogeneous magnetic field \cite{Isaev_17}.
It will be shown below that the results are insensitive to the particular form of repulsive core potential. 
%%%%%%%%%%%%%%%%%%
%%%%%%%%%%%%
\section{Unmagnetized ultracold plasma}
\label{dih_nomag} 
%
%%%%%%%%%%%%%%%%%%
\begin{figure}
  \includegraphics[width = 8.5cm]{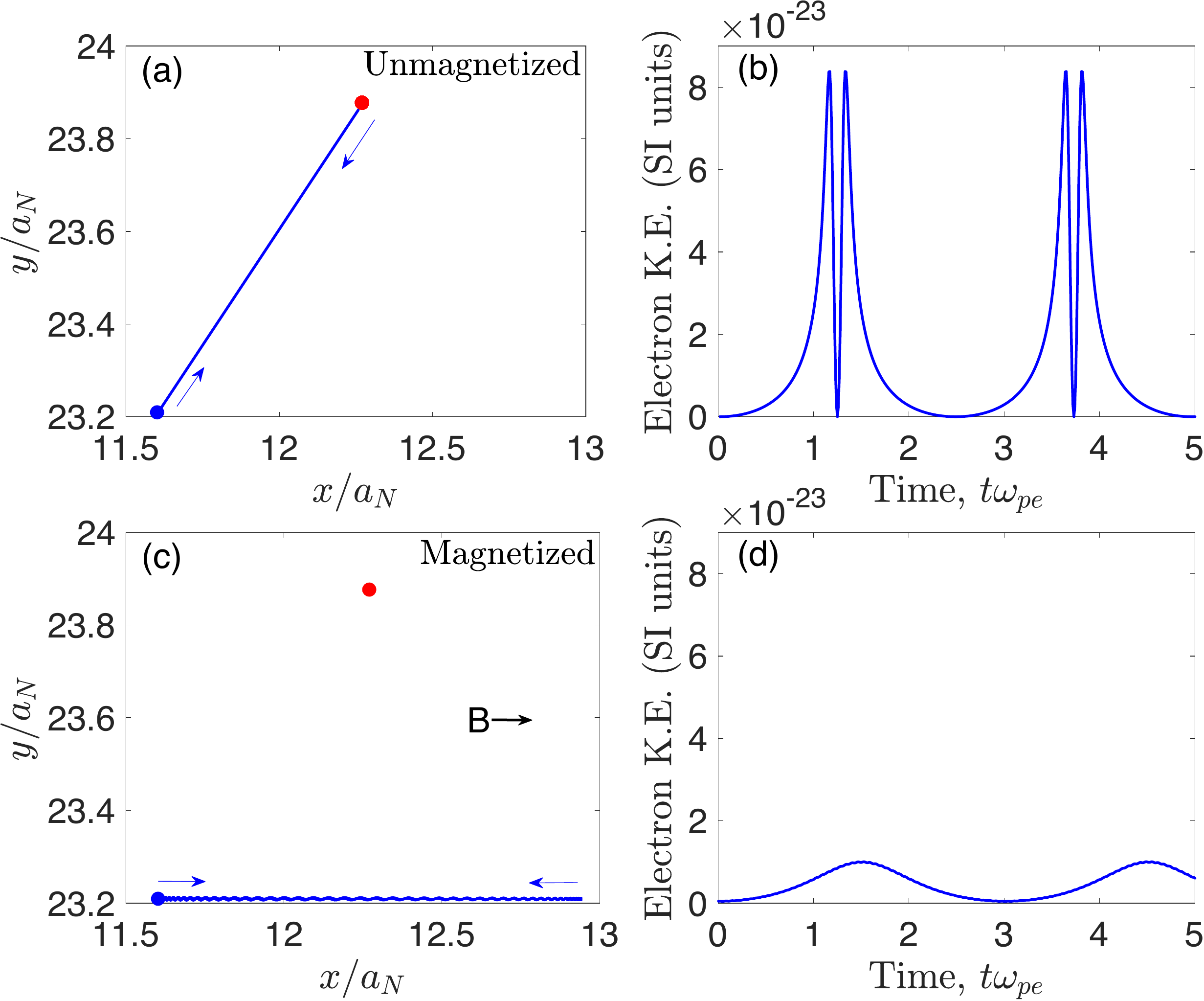}
 \caption{Trajectory (a) and the kinetic energy (b) of an electron moving towards an ion due to an attractive Coulomb potential with a repulsive core, Eq.~(\ref{pot_md}).
 Trajectory (c) and electron kinetic energy (d) of the same interaction in the presence of a magnetic field of $B = 0.32$T. The stationary ion is shown as a red dot, while the blue dot is the initial position of the electron. The normalizing factors $a_N$ and $\omega_{pe}^{-1}$ are $1 \times 10^{-5}$m and $1.7799 \times 10^{-9}$s, respectively.}
 \label{two_part_mag}
\end{figure}
%%%%%%%%%%%%%%%%%%
%%%%%%%%%%%%%%%%%%
\begin{figure*}%[!htb]
 \includegraphics[width = 10cm]{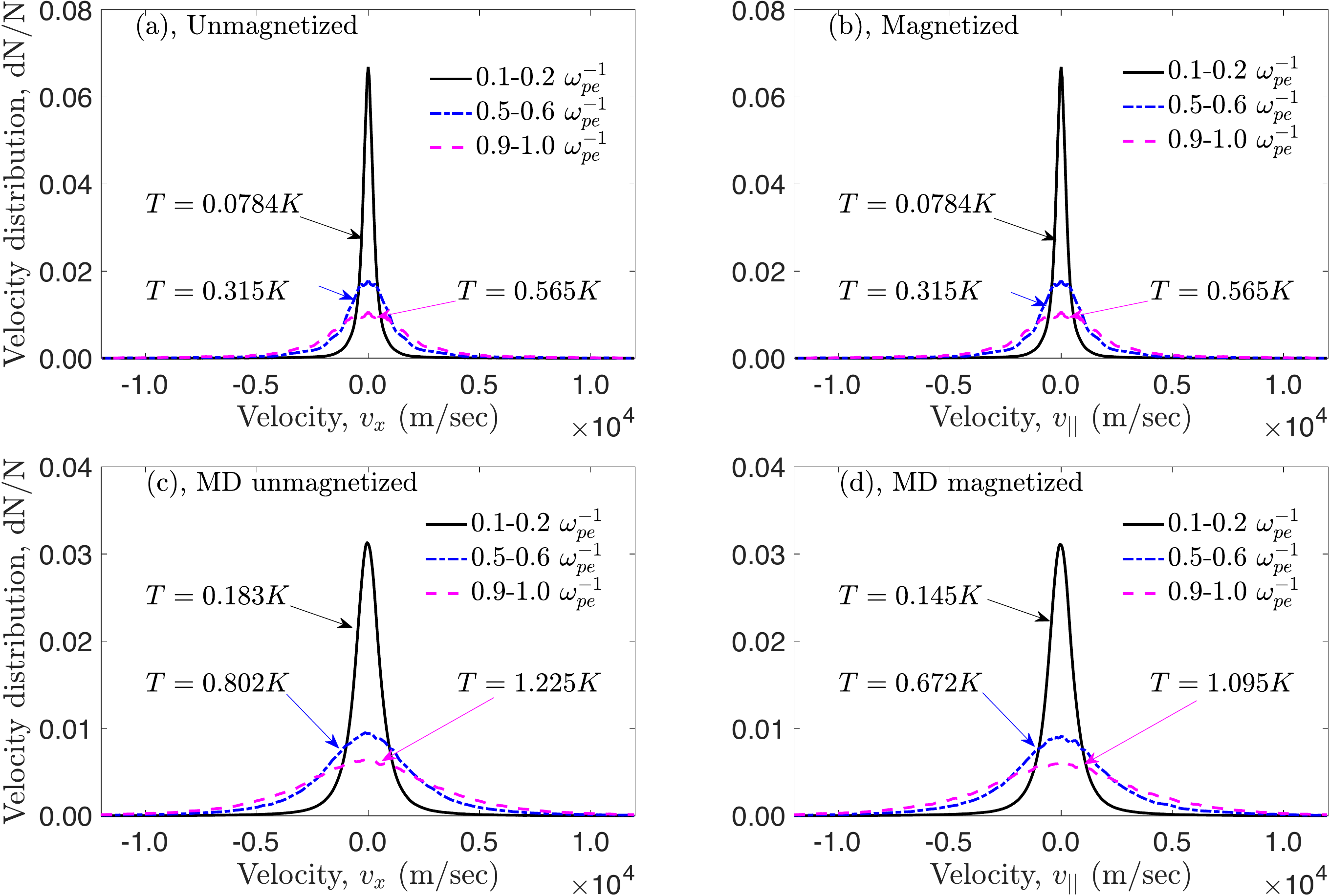}
\caption{Electron velocity distribution for unmagnetized, (a) and (c), and magnetized ($\beta_e = 200$) cases, (b) and (d).  Distributions (c) and (d) were obtained from the MD simulations, while (a) and (b) were from models described in the text.} 
 \label{evel_distb}
\end{figure*}
%%%%%%%%%%%%%%%%%%
\paragraph*{}
In an unmagnetized ultracold plasma, electron-DIH  saturates on a timescale of 1-2 $\omega_{pe}^{-1}$, when the kinetic energy of the electron species becomes comparable to their potential energy at the average interparticle separation, 
%, i.e.  $k_B T_e \sim e^2/a$; %i.e., electron temperature saturation occurs at 
$k_BT_{\textrm{DIH}} \sim e^2/a$ \cite{murillo_01}. 
Figure~\ref{dih_ei} shows the kinetic temperature evolution of electrons (black line) and ions (blue dashed line). 
A rapid increase in the electron temperature due to DIH  occurs immediately, and is saturated within  1-2 $\omega_{pe}^{-1}$. 
A monotonic but slower increase in the electron temperature at later times is attributed to TBR heating. 
The rate of each processes can be determined from the slope of a linear fit in each time interval (dashed line for DIH and dashed-dot line for TBR).
As expected, ion heating is negligible on this short timescale. 

During DIH, electrons gain kinetic energy largely via the ballistic motion associated with their electrostatic attraction to the nearest ion. 
Since the timescale associated with DIH is so short, electrons are not thermalized during this period, and the  electron velocity distribution is not expected to be Maxwellian. Figure~\ref{dih_edistr} shows the electron velocity distribution in the $x$-direction. Black and red solid lines show the distribution of electron velocities at 0.2 $\omega_{pe}^{-1}$ and 1  $\omega_{pe}^{-1}$, respectively. Fits to a Maxwellian, obtained by matching the height of the velocity distribution, are shown as dotted lines. It is clear from the plot that the actual velocity distribution is not Maxwellian. Temperature is not thermodynamic, but rather interpreted in terms of the kinetic temperature of Eqs.~(\ref{temp_prl}) and (\ref{temp_pl}).

\paragraph*{•}
To quantify the electron kinetic energy gained by accelerating towards the nearest ion, we provide a simple model consisting of an immobile ion and an electron with zero initial velocity. The initial distance between the electron and ion was chosen from the initial nearest neighbor electron distribution around ions  from a distribution  where both charged species 
were generated in a random homogeneous manner; as in the MD simulations. The equation of motion $m_e \ddot {\bf r} = - \nabla v_{ei}$ was solved for each individual nearest neighbor pair (picked one pair at a time).
% from the nearest neighbor electron distribution.
The kinetic energy was calculated as the electron moved directly toward the ion over a time interval $\Delta t = \omega_{pe}^{-1}$.  
Figure~\ref{two_part_mag}(a) shows an example electron trajectory, and Fig.~\ref{two_part_mag}(b) the corresponding electron kinetic energy. 
In this example, the initial separation was $r=0.7a$.
After a representative sample of initial nearest neighbor distance configurations was explored, the distribution of electron velocities was obtained.

Figures~\ref{evel_distb}(a) and \ref{evel_distb}(c) show the averaged electron velocity distribution $dN/N$ (probability of finding electrons in velocity range of $v_x$ to $v_x + 50$ m/s) obtained from the model and from MD simulations, respectively. The velocity distributions of the model and simulations qualitatively agree. %, and neither is Maxwellian. 
The average kinetic temperature over three separate time intervals ($0.1-0.2$, $0.5-0.6$ and $0.9-1.0 \omega_{pe}^{-1}$), calculated using Eq.~(\ref{temp_prl}), are also indicated in the figure. While temperatures obtained from the model were $0.0784$K, $0.315$K and $0.565$K (i.e. $1/\Gamma_e = 0.0628,  0.2522$  and  $0.4523$ respectively), the temperatures obtained from the MD simulations were approximately twice as high with values $0.183$K, $0.802$K and $1.225$K (i.e. $1/\Gamma_e = 0.1465, 0.6421$    and $0.9807$ respectively). 
%\textcolor{blue}{Though the analytic model provides a qualitative picture of the origin of the kinetic energy of each electron in the ultracold plasma system, the temperature obtained using this two particle model is almost half of the temperature gained by electrons in the full MD simulation. 
%We found that the reason lies in the multi-body interactions in the plasma. 
The additional heating observed in the MD simulations, compared to the binary interaction model, must be associated with many-body interactions. 
To test this, we carried out two particle MD simulations for several pairs of nearest neighbors and compared the kinetic energy gained by the electron with its value from the full MD simulation in presence of all other charges. In each case, we found that the kinetic energy obtained by electron in full MD is greater than what it attains in two particle simulation. 

%%%%%%%%%%%%%%%%%%
\begin{figure*}[!htb]
 \includegraphics[width = 10.0cm]{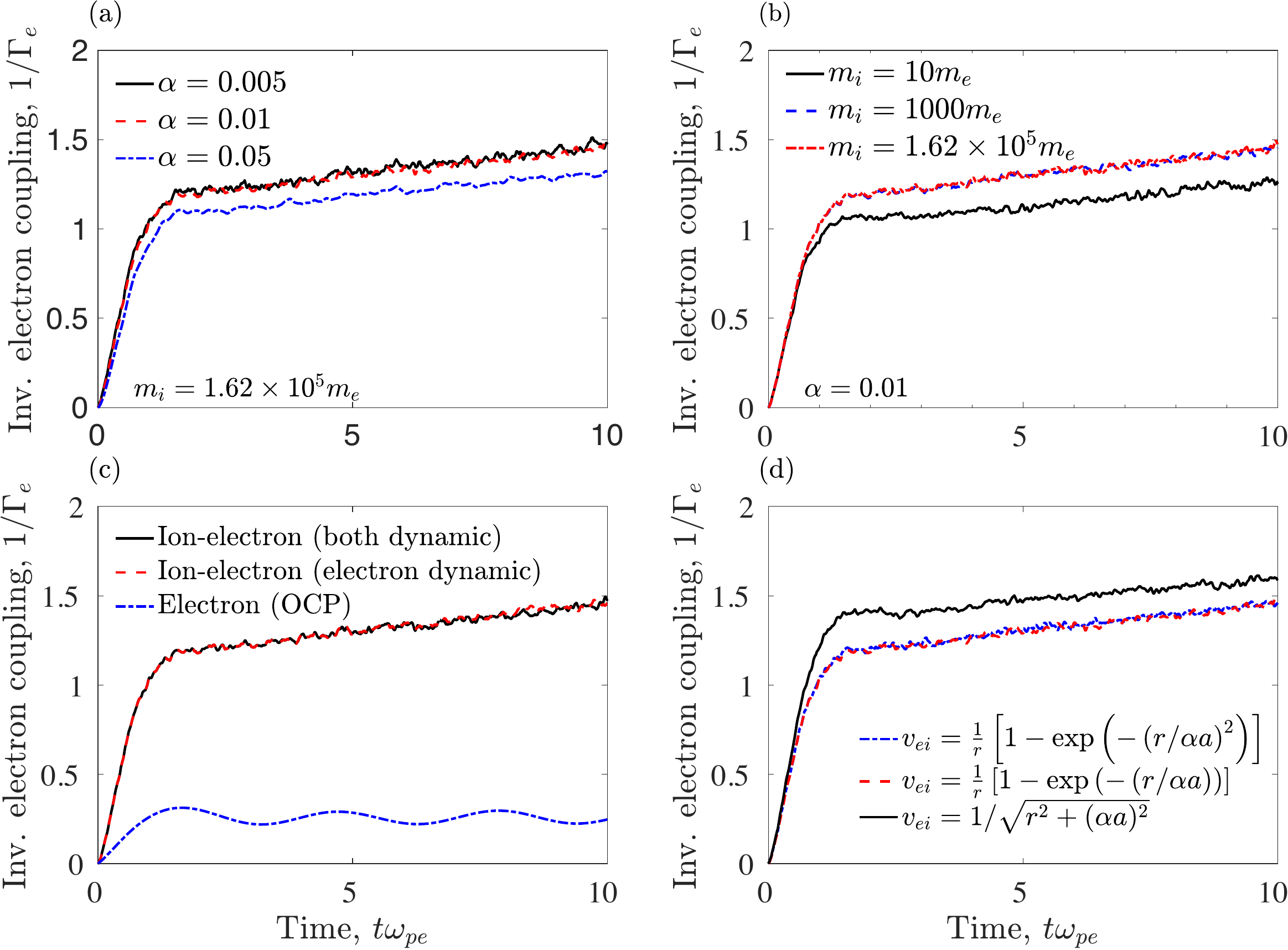}
\caption{(a) Electron temperature profile dependence on repulsive core scale $\alpha$, and (b) the ion to electron mass ratio $m_i/m_e$. (c) Comparison of electron temperature profiles when both species are mobile, electrons are mobile and ions are immobile, and 
electrons are the only species in the simulation (OCP). (d) Compares electron heating for three different models of electron-ion interaction potential.} 
 \label{dih_malp}
\end{figure*}
%%%%%%%%%%%%%%%%%%
Figure~\ref{dih_malp} presents the dependence of the electron temperature evolution profile on (a) the repulsive core parameter $\alpha$, (b) the ion to electron mass ratio $m_i/m_e$, (c) the choice of simulation models and (d) the choice of pseudo-potential for electron-ion interaction.
Figure~\ref{dih_malp}a suggests that the $T_e$ evolution profile does not appreciably change for values of the repulsive core parameter $\alpha<0.01$ over the time interval of interest. 
Here, close interactions $r < \alpha a$ are sufficiently rare that further decreasing the value of  $\alpha$ causes negligible changes to the heating. 
For the remainder of this work, we choose the value $\alpha=0.01$.  
Figure~\ref{dih_malp}b shows 
% the dependence of the $T_e$ evolution profile on the ion to electron mass ratio for $m_i/m_e = 10, 1000$ and $1.62 \times 10^5$. It is evident 
that the $T_e$ profiles are indistinguishable for the mass ratio of $1000$ and $1.62 \times 10^5$. This suggests that for any realistic mass ratio  $m_i/m_e \geq 1000$, electron and ion dynamical  timescales are sufficiently well separated that ion dynamics does not affect the electron heating process.  For the remainder of this work we choose  $m_i/m_e = 1.62 \times 10^5$, which corresponds to strontium plasma. 

Figure~\ref{dih_malp}c provides a comparison of $T_e$ evolution profiles obtained from the one-component plasma (OCP) model (blue dashed line) with those obtained from two-component plasma (TCP) models. The black solid  line represents the $T_e$ profile when both (electron and ion) species were dynamic during the simulation. The red dashed line  represents $T_e$ profile when only electrons were dynamic and ions were held stationary. Both TCP models provide the same $T_e$ evolution profile, suggesting that ion motion does not influence electron heating over this time scale. This agrees with earlier simulations made by  Kuzmin et al \cite{kuzmin_prl_02}.  
An OCP model predicts a much lower $T_{\textrm{DIH}}$ and suggests that the electrons could remain strongly coupled ($\Gamma_e > 1$) after the saturation of DIH. 
The OCP model does not agree with the TCP models or with experimental observations where electrons are observed to be weakly coupled \cite{roberts_04}. 
We expect the reason is that the early stage ($\Delta t < \omega_{pe}^{-1}$) heating of electrons is dominated by kinetic energy gain by attraction towards
nearest neighbor ions, which is absent in the OCP DIH models \cite{murillo_01}. The Yukawa-OCP model has been found to accurately explain ion temperature evolution, in agreement with experiments \cite{McQuillen_15}. This is likely because at late times ($\Delta t \sim \omega_{pe}^{-1}$) electrons have already
 moved towards ions, so that ions move in presence of a polarized (screened) charge cloud created by electrons (i.e, a YOCP).

To investigate the role of  pseudo-potential choice on the electron heating mechanism, we modeled the electron-ion interaction using three different forms:
$v_{ei}=e^2/\sqrt{r^2 + (\alpha a)^2}$ (O' Neil's form \cite{kuzmin_prl_02}),  $v_{ei}=(e^2/r)(1 - \exp{(-r/\alpha a))}$ (Hansen' form \cite{hansen_prl_78}) and the Kelbg form (Eq.~\ref{pot_md}) \cite{kelbg_63}. Figure~\ref{dih_malp}d shows that the $T_e$ profiles obtained from the simulations using all three different forms of pseudo-potential are in good agreement for a small value of the repulsive core parameter $\alpha = 0.01$. 

%%%%%%%%%%%%
\section{Magnetized ultracold plasma}
\label{dih_mag} 
\paragraph*{}
When an external magnetic field is applied, electrons gyrate around the field lines. 
%As the magnetic field strength increases,  the gyroradius  decreases  and when it is sufficiently high, 
At high field strength the gyroradius is so small that the electrons move primarily in one-dimension, parallel to the magnetic field, and consequently  
%When electrons are constrained to nearly one dimension, they 
gain kinetic energy in only this direction.
%, leading to DIH in the parallel direction only. 
Because electron kinetic energy gain is exceptionally slow in the cross field direction, this leads to an effective drop in the total kinetic $T_e$ in comparison to an unmagnetized plasma. 

%
%%%%%%%%%%%%%%%%%%
\begin{figure}
  \includegraphics[width = 8cm]{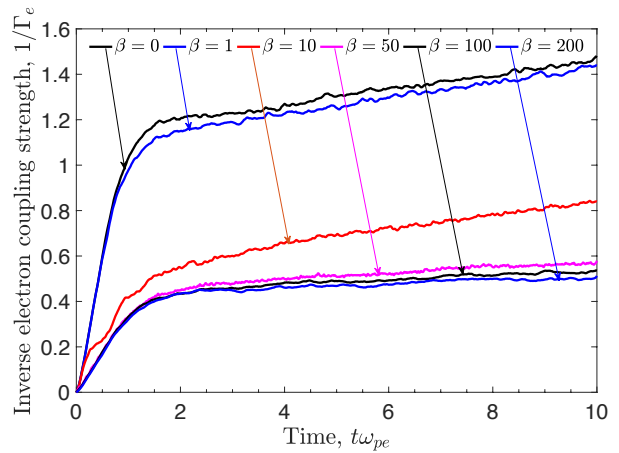}
 \caption{Electron temperature evolution at different  magnetic field strengths. 
  The repulsive core parameter $\alpha = 0.01$ and the ion mass corresponds to the Strontium ion.}
 \label{dih_mageff}
\end{figure}
%%%%%%%%%%%%%%%%%%
%%%%%%%%%%%%%%%%%%
\begin{figure}[!htb]
 \includegraphics[width = 8cm]{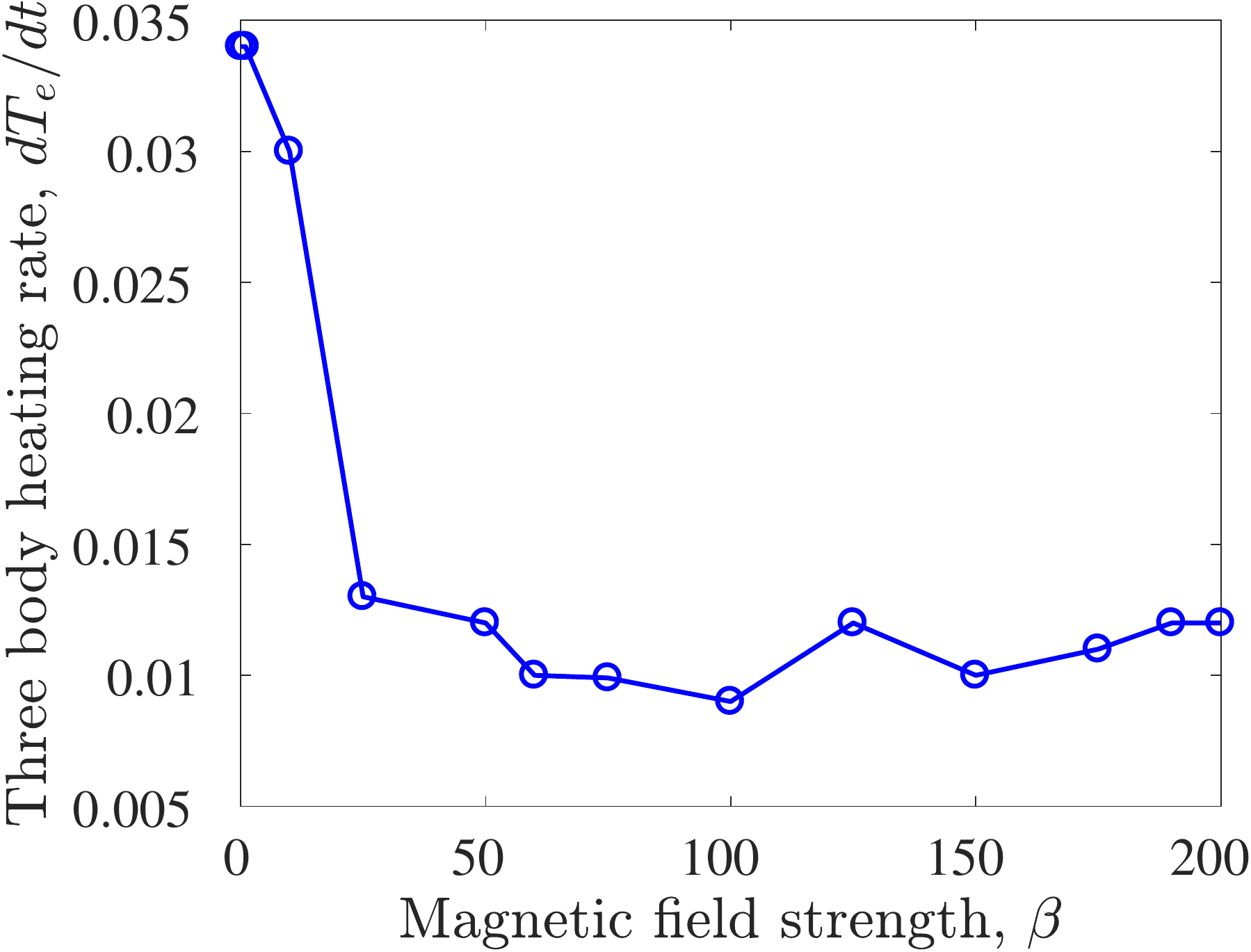}
 \caption{TBR electron heating rate for different magnetic field strengths. The repulsive core parameter $\alpha = 0.01$ and ion mass corresponds to strontium ion.}
 \label{tbr_mageff}
\end{figure}
%%%%%%%%%%%%%%%%%%

Figure~\ref{dih_mageff} shows the effect of increasing magnetic field strength in the MD simulations. At a sufficiently strong magnetic field, the total kinetic temperature due to the DIH  saturates (in $\sim \omega_{pe}^{-1}$) at approximately one third of the value in the case of no magnetic field. We further observe a drop in TBR heating rates at later times with the increase in magnetic field strength.  Figure~\ref{tbr_mageff} shows that the TBR heating rate saturates at   one-third of its value in the unmagnetized case. The TBR heating rate is calculated as the slope of the $T_e$ profile in the time interval $1.5 \omega_{pe}^{-1} < t < 10 \omega_{pe}^{-1}$ (see Fig.~\ref{dih_ei}).
%%%%%%%%%%%%%%%%%%
\begin{figure}[!htb]
 \includegraphics[width = 8cm]{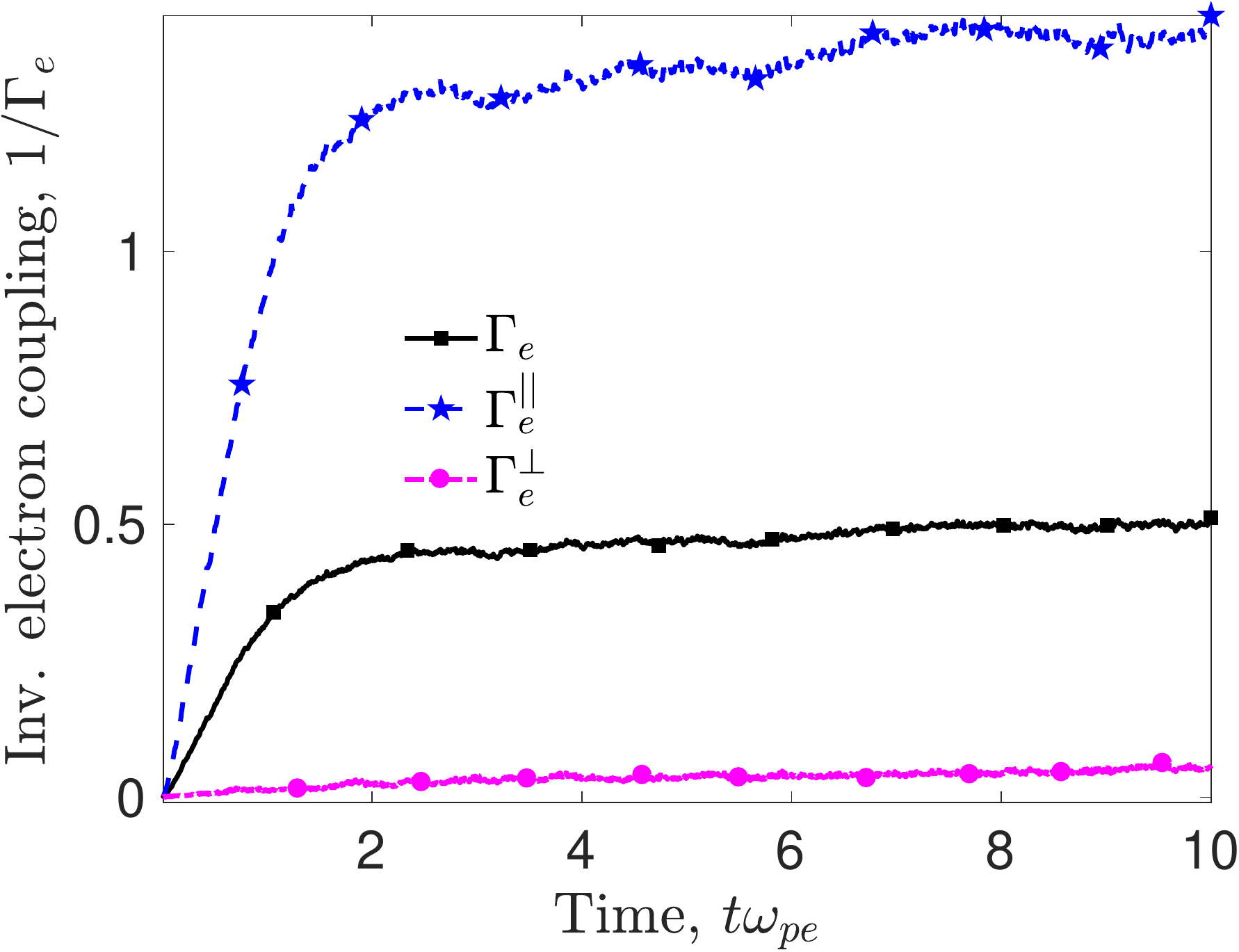}
 \caption{Electron temperature evolution in directions parallel and perpendicular to the magnetic field. The repulsive core and magnetic field strength
 are $\alpha = 0.01$ and $\beta_e = 200$.}
 \label{dih_mag_Tprpl}
\end{figure}
%%%%%%%%%%%%%%%%%%
%%%%%%%%%%%%%%%%%%
\begin{figure*}
 \includegraphics[width = 0.98\textwidth]{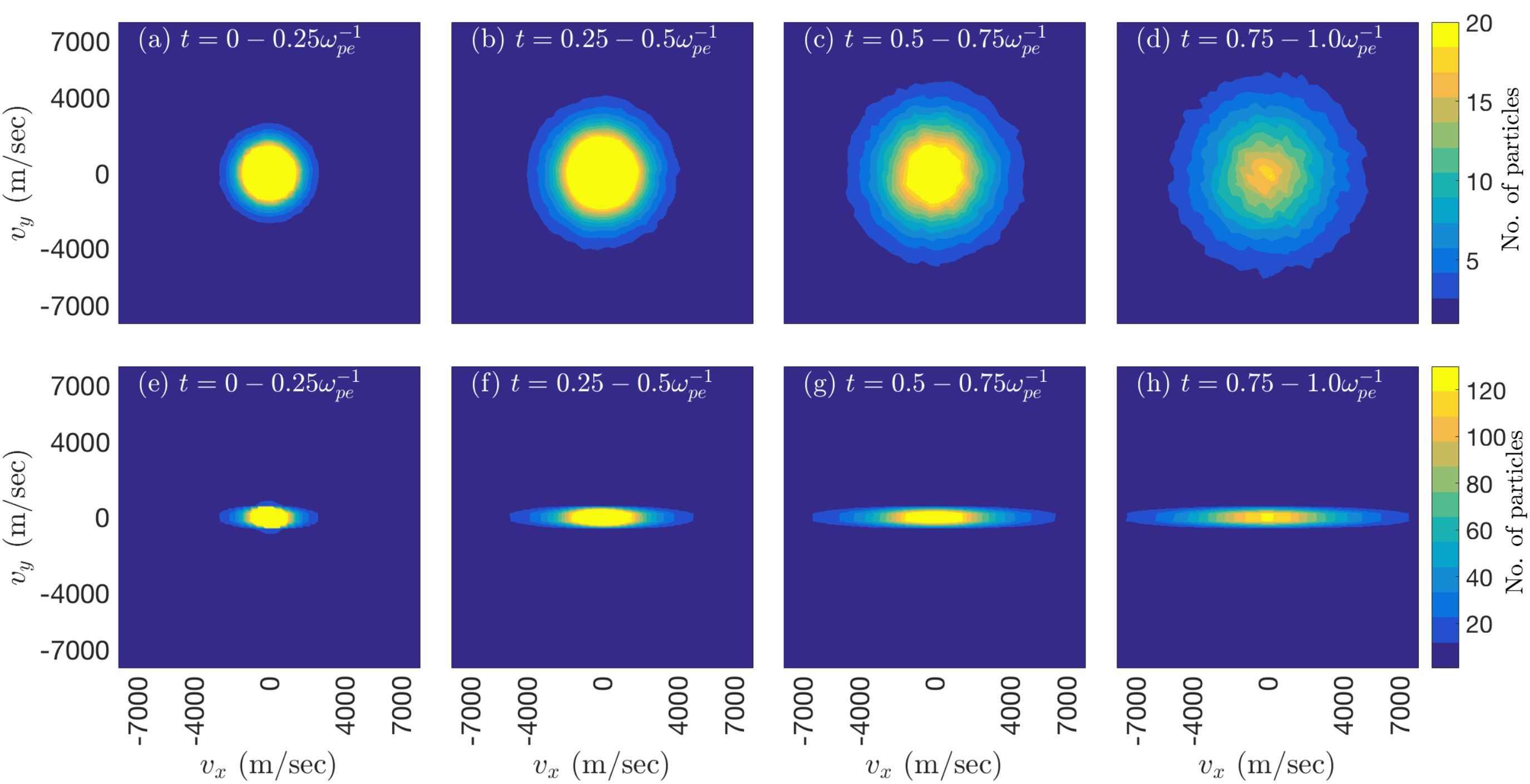}
 \caption{Two dimensional velocity distribution function (parallel (x-component) and perpendicular (y-component)) electron in (a)-(d) an unmagnetized and (e)-(h)
 magnetized ($\beta_e = 200$) ultracold plasma. %Velocities of particles correspond to a time of 3 electron plasma periods. 
 }
 \label{vel_dist_mag_umag}
\end{figure*}
%%%%%%%%%%%%%%%%%%
\paragraph*{}
Due to the strong magnetic field, a strong temperature anisotropy  develops in the plasma \cite{ott_pre_17a}. 
Figure~\ref{dih_mag_Tprpl} shows the evolution of the kinetic electron temperature in the presence of  a strong magnetic field ($\beta=200$). It is evident  that the parallel kinetic temperature (blue dashed line) rises as it does  in an unmagnetized plasma, and saturates once the kinetic energy of the electrons in this direction becomes comparable to the potential energy of particles at an average interparticle separation i.e, $k_B T_{||}  \sim e^2/a$.  

However, because of the strong magnetic field, the heating time in the perpendicular direction to the field becomes so long that there 
is  essentially no increase in the kinetic temperature in this direction. The magenta dash-dotted line in the Fig.~\ref{dih_mag_Tprpl} shows the evolution of the perpendicular  kinetic temperature. 
The black line in Fig.~\ref{dih_mag_Tprpl} shows the total kinetic temperature from Eq.~(\ref{temp_tot}). 
Out of three directions, the temperature remains negligible in two of them due to the strong magnetic field. This leads to a total electron temperature  which is close to one-third of electron temperature in the unmagnetized case. Typically, the ratio $T_{\parallel}/T_{\perp} \approx 30$ at 10$\omega_{pe}^{-1}$ with
the magnetic field strength of $\beta=200$. 

The temperature anisotropy can also be observed in the velocity distribution function; see Fig.~\ref{vel_dist_mag_umag}. 
%, shows the evolution of the electron velocity distribution function  for magnetized and unmagnetized cases in  two dimensional velocity space. 
Subplots (a) to (d) show the averaged electron velocity distribution over four time intervals in the range from 0 to 1 $\omega_{pe}^{-1}$. A spherically symmetric (circularly symmetric in the 2D plot) electron velocity distribution is observed when there is no magnetic field. In contrast, the electron velocity distribution is highly asymmetric in the presence of a strong applied magnetic field (subplots (e) to (h)). 

The spontaneously generated temperature anisotropy, will relax at a rate that depends on the coupling strength $\Gamma$ and the magnetic field strength $\beta$ \cite{dubin_05,Anderegg_09,ott_pre_17a}. 
Recent results \cite{baalrud_17} (for the one component plasma) suggest that for a coupling strength of $\Gamma = 0.1$ and  magnetic field strength of $\beta = 100$, the temperature isotropization time is $\sim 10^6 \omega_{pe}^{-1}$. 
This would be a sufficiently long delay that the electron temperature anisotropy  would last for the complete ultracold plasma lifetime, which can last up to 250 $\mu s$ (or few hundreds of $\omega_{pi}^{-1}$).

Similar to the unmagnetized case, the early time ($\Delta t < \omega_{pe}^{-1}$) kinetic energy gained by electrons is primarily due to their ballistic motion in the presence of a nearest-neighbor ion (with approximately half of the contribution being due to the multi-body interactions).
%However, because of the magnetic field, 
At high field strength, the ballistic motion is not directly toward the ion, but rather is restricted to the direction parallel to the magnetic field. 
This effectively reduces the average electrostatic potential energy accessible to the electrons in comparison to an unmagnetized plasma. 

To demonstrate this, we again introduce a simple model. 
The equation of motion $m_e d \mathbf{v}_e/dt = - \nabla v_{ei} + \mathbf{v}_e \times \mathbf{B}$ was solved for each individual nearest neighbor pair (picked one at a time) from the nearest neighbor electron distribution (as described in the unmagnetized case) and the electron kinetic energy was calculated. 
%over $\Delta t = \omega_{pe}^{-1}$. 
%Once the magnetic field is applied in the x-direction, electron motion constrained in that direction and gains no kinetic energy in perpendicular direction to the field 
An example trajectory is shown in Fig.~\ref{two_part_mag}(c), and the corresponding kinetic energy in Fig.~\ref{two_part_mag}(d).
%This further leads to drop in the total kinetic energy of electron (Fig.~\ref{two_part_mag}(d)).
Similar to the unmagnetized case, we use this model for all possible nearest neighbor distances between electrons and ions obtained from a random initial distribution. 
%\textcolor{red}{Here, can you show that the KE in the magnetized case is the same as if you take just the x-component of the unmagnetized case?} 
 %Each time, the electron motion is evolved for $1 \omega_{pe}^{-1}$ and its velocity is collected in the parallel direction to the magnetic field. 
Figure~\ref{evel_distb}(b) shows that the electron velocity distribution in the parallel direction obtained from this model qualitatively matches that one obtained from the MD simulations with $\beta_e = 200$ (Fig.~\ref{evel_distb}(d)). The time averaged (over $0.1 \omega_{pe}^{-1}$) velocity distributions obtained from both ways do not follow the Maxwellian.
The average parallel kinetic temperature in the time intervals $0.1-0.2$, $0.5-0.6$ and $0.9-1.0 \omega_{pe}^{-1}$ from the model are
$0.0784$K, $0.315$K and $0.565$K (i.e. $1/\Gamma_e=0.0628, 0.2522$ and  $0.4523$), respectively (same as the unmagnetized case because there is no force due to the magnetic field in the parallel direction). 
%\textcolor{red}{this is confusing since you defined the kinetic temperature to be 3D, and this is citing only the 1D energy, yet you are just calling it the ``kinetic temperature''. I also do not understand exactly how the initial distribution is set up here. Are you taking the starting positions sampled throughout 3D?}
The average parallel temperatures obtained using MD simulations during these same intervals are $0.145$K, $0.672$K and $1.095$K (i.e. $1/\Gamma_e = 0.1161, 0.5380$  and $ 0.8766$).
%\textcolor{red}{Again, confusing that you don't call this parallel energy. Isn't the point of the paper that these should be 1/3 the value?}

\begin{figure}
 \includegraphics[width = 8cm]{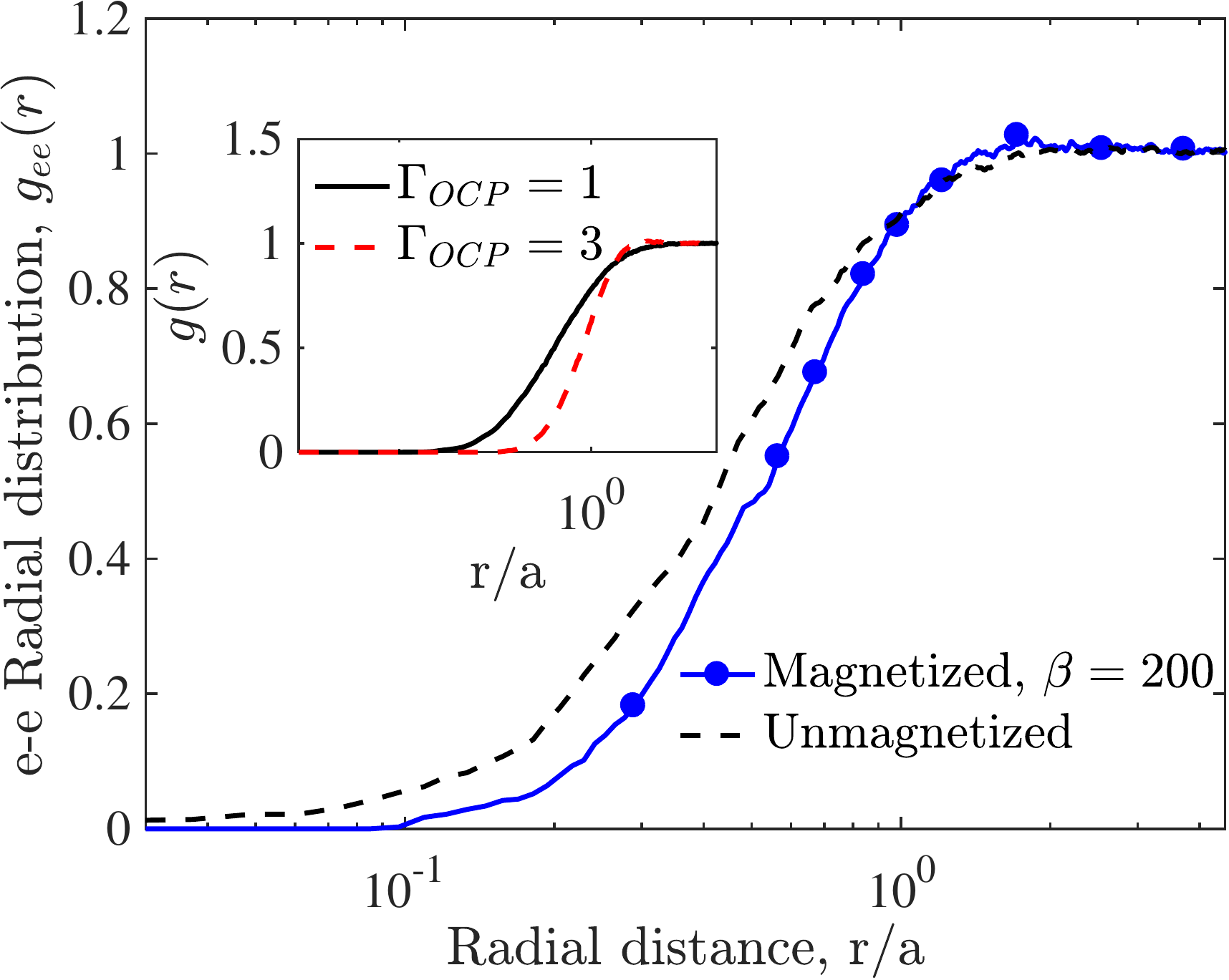}
 \caption{Electron radial distribution (averaged over 3-5 $\omega_{pe}^{-1}$) for magnetized (full blue line) and unmagnetized (dashed black line) cases.
 Inset shows RDFs for coupling strength $\Gamma = 1$ and $3$ for an equilibrium OCP system.}
 \label{eerdf}
\end{figure}

Finally, as a measure of increased coupling strength, we calculate the radial distribution function (RDF) for the electrons, which describes the density surrounding a fixed particle referenced to the background. In a homogeneous and isotropic plasma, this depends only on the radial distance from the reference particle. 
In molecular dynamics, the RDF is calculated by binning distances between all the particles to determine how many particles lie within distance $r$ to $r+dr$
away from the reference particle. Further, these values are normalized by the particle histogram for an ideal gas which has no correlations. 
In Fig.~\ref{eerdf}, we plot RDFs  of electrons  for an unmagnetized plasma (dashed black line) and for a plasma with magnetic field of strength $\beta_e =200$ (blue solid line). Each RDF is averaged over duration $3-5 \omega_{pe}^{-1}$, a timescale at which electron DIH is saturated. 
The fall of the RDF towards $r \rightarrow 0$ is steeper in the case with a magnetic than in the case without.
This steepness  is a signature of a higher coupling strength. This fact can be verified by RDFs obtained from the OCP for $\Gamma = 1$ and $\Gamma = 3$;  see the inset of Fig.~\ref{eerdf}.
%
%%%%%%%%%%%%%%%%%%
%%%%%%%%%%%%%%%%%%
%
\section{Possible implications for experiments}
\label{ucp_impl}
\subsubsection{1-10 $\omega_{pe}^{-1}$ (ns) timescale}
Experiments focused on  electron dynamics make measurements on 1-200 ns timescales  \cite{Chen_pop_16,Chen_17}. Such experiments are designed in such a way that the electrons possess a very small initial kinetic energy due to the photoionization of ultracold atoms \cite{Chen_pop_16}. Here, we suggest that an external magnetic field of approximately one-tenth of a Tesla or higher could reduce the subsequent DIH and TBR heating to one-third of the value in the absence of a magnetic field. 
This will increase the effective electron coupling strength by a factor of three. Recent experiments (at electron time scales (ns)) have reported an electron coupling strength $\Gamma_e \sim 0.35$ ($T_e \sim 1.6$K) \cite{Chen_17}, so the presence of an external magnetic field creating $\beta_e \sim 50$ (or higher) will be sufficient to observe the $\Gamma_e \sim 1$ on this timescale.

\subsubsection{1 - 100 $\omega_{pi}^{-1}$ ($\mu$s) timescale}
Experiments focused on ion dynamics make  measurements at  $\mu$s timescales. Over this period ($> 10^3 \omega_{pe}^{-1}$), electron species continuously gain  kinetic energy due to the TBR heating process. A magnetic field of  $\beta_e \sim 50$ (0.16 T for $n_e =1\times 10^{14}$m$^{-3}$) is sufficient to  reduce the DIH and TBR heating by a factor of three over a timescale of several $\omega_{pe}^{-1}$, but  to extend any relaxation of the temperature anisotropy to $\mu$s timescales, experiments will need a strong magnetic field satisfying $\beta_e > 100$. The sustained temperature anisotropy may extend the increased electron coupling strength (compared to unmagnetized case) to the $\mu$s timescale.

Though UCPs develop a strong temperature anisotropy due to the applied magnetic field, such plasmas are not expected to be susceptible to 
temperature or pressure anisotropy driven instabilities (such as fire-hose \cite{Chandrasekhar435,rosenbluth1956} or Weibel \cite{weibel_59}). 
These instabilities are known to exist at high values of plasma beta ($\beta_{pl} = n k_B T/(B^2/2\mu_0) \gtrsim 1$). 
The $\beta_{pl}$ is related to the magnetic field strength as $\beta_{pl} = (\tilde{v}/c)^2/\beta^2$. 
For UCPs, temperature is a very small quantity, $\tilde{v}/c << 1$ ($\tilde{v}$ is the mean thermal velocity). 
Also, we are studying UCPs under strong magnetic field strengths ($\beta > 100$). Under these conditions,  $\beta_{pl}$ becomes a very small value.

%
%%%%%%%%%%%%%%%%%%%
%%%%%%%%%%%%%%%%%%%
\section{Summary}
\label{summ}
In summary, we addressed the issue of the electron heating and its reduction through the application of an external magnetic field in an electron-ion ultracold plasma. Using classical MD simulations, we showed that in the presence of a strong magnetic field ($\beta_e \geq 50$), electron DIH 
is reduced by a factor of three compared to its value in an unmagnetized plasma.  We also showed that the electron TBR heating rate was reduced  by the factor of three at a similarly strong magnetic field strength.  The reduction in heating occurs due to electron motion being constrained to one-dimension (parallel to magnetic field) resulting in little heating in the perpendicular direction.  
 The  ultracold plasma parameters suggested that a magnetic field of approximately a Tesla would be sufficient to see this effect  in  experiments at timescales
 of the lifetime of UCP.  This suggests the possibility to observe an increased effective coupling strength of electrons along with the moderately coupled ions in UCP experiments.  We also see that a strong external magnetic field causes a strong temperature anisotropy $T_{\parallel}/T_{\perp} \sim 30$ in ultracold plasmas  during their evolution.  
Using a simple model consisting of individual electron-ion pairs, it was shown that approximately one-half of the electron disorder induced heating occurs due to electron motion directly toward its nearest neighbor ion at the early times ($0 < t < 1 \omega_{pe}^{-1}$).  
%%%%%%%%%%%%%%%%%%%
%%%%%%%%%%%%%%%%%%%
\begin{acknowledgments}

\textcolor{black}{We thank N. Shaffer for useful discussions. This material is based upon work supported by the Air Force Office of Scientific Research under award number FA9550-16-1-0221.}
It used the Extreme Science and Engineering Discovery Environment (XSEDE) \cite{towns_xsede}, which is supported by NSF grant number ACI-1053575, under project award No.~PHY-150018. This research was supported in part through computational resources provided by The University of Iowa.

\end{acknowledgments}

%%%%%%%%%%%%%%%%%%%%%%%%%
%~~~~~~~~~~~~~~~~~~~~~~~~~~~~~~~~~~~~~~~~~~~~~~~~
%\bibliographystyle{unsrt}
\bibliography{thermo_ei}
%
%%%%%%%%%%%%%%%%%%%%%%%%%%%%%%%%%
%~~~~~~~~~~~~~~~~~~~~~~~~~~~~~~~~~~~~~~~~~~~~~~~~~~~~~~~~~~~~~~~~~~~~~~~~~~~~~~~~~~~~~~~~~~~~~~~~~
\end{document}